 \documentclass[pmlr,twocolumn,10pt]{jmlr} 

\usepackage{array}
\usepackage{multirow}
\usepackage{graphicx}
\usepackage{booktabs}
\usepackage{siunitx}
\usepackage{xcolor}
\usepackage{easyReview}
\renewcommand{\add}[1]{{\color{black}#1}}

\newcommand{\equal}[1]{{\hypersetup{linkcolor=black}\thanks{#1}}}

\theorembodyfont{\upshape}
\theoremheaderfont{\scshape}
\theorempostheader{:}
\theoremsep{\newline}

\jmlrvolume{LEAVE UNSET}
\jmlryear{2023}
\jmlrsubmitted{LEAVE UNSET}
\jmlrpublished{LEAVE UNSET}
\jmlrworkshop{Machine Learning for Health (ML4H) 2023} 

 \title[Point-of-Care Real-Time Signal Quality for Fetal Doppler Ultrasound]{Point-of-Care Real-Time Signal Quality for Fetal Doppler Ultrasound Using a Deep Learning Approach}

\author{%
\Name{Mohsen Motie-Shirazi} \Email{mohsen.shirazi@dbmi.emory.edu}\\
\addr Emory University, USA
\AND
\Name{Reza Sameni}
\Email{rsameni@dbmi.emory.edu}\\
\addr Emory University, USA
\AND 
\Name{Peter Rohloff}
\Email{peter@wuqukawoq.org}\\
\addr Center for Indigenous Health Research, Wuqu' Kawoq | Maya Health Alliance, Guatemala
\AND
\Name{Nasim Katebi}\equal{Joint Senior Authors} 
\Email{nkatebi@emory.edu}\\
\addr Emory University, USA
\AND
\Name{Gari D. Clifford}\footnotemark[1] \Email{gari@gatech.edu}\\
\addr Emory University \& Georgia Institute of Technology, USA
}
\begin{document}

\maketitle

\begin{abstract}

In this study, we present a deep learning framework designed to integrate with our previously developed system that facilitates large-scale 1D fetal Doppler data collection, aiming to enhance data quality. This system, tailored for traditional Indigenous midwives in low-resource communities, leverages a cost-effective Android phone to improve the quality of recorded signals. We have shown that the Doppler data can be used to identify fetal growth restriction, hypertension, and other concerning issues during pregnancy. However, the quality of the signal is dependent on many factors, including radio frequency interference, position of the fetus, maternal body habitus, and usage of the Doppler by the birth attendants. In order to provide instant feedback to allow correction of the data at source, a signal quality metric is required that can run in real-time on the mobile phone.

In this study, 191 DUS signals with durations mainly in the range between 5 to 10 minutes were evaluated for quality and classified into five categories: Good, Poor, (Radiofrequency) Interference, Talking, and Silent, at a resolution of 3.75 seconds. A deep neural network was trained on each 3.75-second segment from these recordings and validated using five-fold cross-validation.

An average micro F1 = 97.4\% and macro F1 = 94.2\% were achieved, with F1 = 99.2\% for `Good' quality data. These results indicate that the algorithm, which will now be implemented in the midwives' app, should allow a significant increase in the quality of data at the time of capture.
\end{abstract}

\section{Introduction}
While global medical advancements have significantly decreased mortality rates in many countries, childbirth remains perilous for both mother and child. Maternal mortality is as high as 1000 per 100,000 live births in some countries and regions,  Annually, there are nearly 2 million stillbirths and 2.3 million children die in the first month of life \citep{who20,unicef23}, predominantly in low and middle-income countries (LMICs). Notably, LMICs contribute to around 98\% of these perinatal deaths, which can be attributed to factors such as the dearth of trained medical professionals, high medical equipment costs, and the subsequent limited access to specialized prenatal care, especially in remote areas \citep{zupan05}. Due to the COVID-19 pandemic and its aftermath, perinatal statistics have worsened in many settings \citep{Tannep659}. 

Fetal Growth Restriction (FGR) and congenital abnormalities are significant contributors to these mortality rates \citep{van11,lawn16}. Proactive identification and prompt intervention for such issues can greatly reduce the risks of neonatal morbidity and mortality. In high-income countries, ultrasound imaging, especially fetal echocardiography for diagnosis of FGR, is the standard method for fetal health monitoring. However, high device costs (from \$2,000 to \$10,000) and the requirement for specialized training, hinder their widespread adoption in LMICs \citep{who15}. 

 Over the last 10 years, we have developed and implemented a cost-effective perinatal screening system, which leverages low-cost smartphones to collect key data during routine wellness visits \citep{martinez17,stroux16}. The system was co-designed with traditional Indigenous midwives (TIMs) in rural Guatemala who serve a Maya population with a
 history of racial/ethnic discrimination and a consequent lack of trust in medical (and other) authorities \citep{schooley2009factors}. Comprising a  low-cost (\$10) one-dimensional Doppler ultrasound (1D-DUS) transducer connected to smartphones via a standard audio cable, it enables the monitoring of fetal cardiac activity, a critical measure in identifying abnormal fetal conditions \citep{sandmire98}. In previous work, we demonstrated that the 1D-DUS signals can be used to accurately estimate both the fetal heart rate and gestational age \citep{valderrama19,katebi23}. Such estimates can lead to identification of fetal cardiac abnormalities and FGR. Nevertheless, the quality of the input signal is essential in ensuring accurate assessments of physiology, and must be implemented before any system can be deployed in the hands of health practitioners.

Historically, the quality of the recorded signals, even by trained TIMs, has posed challenges. Notably, it was determined that approximately 40\% of the recordings made by TIMs were of low quality \citep{martinez17}. The diminished signal quality can be attributed to diverse types of noise and interference. Common issues include incorrect cable connections that capture ambient sounds from sources like human conversations and animal noises or that lead to recordings that are faint or completely silent. Moreover, interference from electronic devices, particularly mobile phones, presents another significant challenge. Most of these issues cannot be easily addressed by post-processing solutions, highlighting the critical need for a system that delivers real-time feedback and recommendations during the data collection process.


Efforts to classify the quality of DUS signals and distinguish high-quality data have been ongoing. Notably, high-quality DUS signals often exhibit a consistent relative periodicity of fetal heartbeats. This characteristic has prompted past research to suggest DUS quality evaluations based on the repeated presence of these signal patterns. In these works, essential signal features are typically extracted and fed into machine learning algorithms for signal classification. In one of the earliest studies, a Signal Quality Index (SQI) was introduced, utilizing sample entropy and the relative wavelet energy distribution within a logistic regression model framework \citep{stroux15}. This approach enabled differentiation between high- and low-quality signals (as determined by expert consensus), achieving an impressive accuracy of 95\%. In a subsequent study, four SQIs were extracted for each fetal heartbeat segment from 1D-DUS signals collected from 17 subjects \citep{Valderrama17}. These SQIs were then integrated with sample entropy and a power spectrum density ratio, achieving an 85\% accuracy rate in classifying expert-annotated data in a 5-class task. More recently, a two-step classifier that combined logistic regression and a multiclass support vector machine (SVM) was utilized to analyze 195 fetal recordings \citep{valderrama18}. Recordings were split into 0.75-second segments and classified by three annotators into five quality categories. From five such segments, 3.75-second windows were created. For each, 88 features were extracted and later refined to 17 via feature selection. Consequently, the study achieved impressive micro-averaged and macro-averaged F1 scores of 96.0\% and 94.5\%, respectively. 

While prior studies have shown remarkable achievements in DUS signal quality classification, relying predominantly on handcrafted features and conventional machine learning techniques, there is potential to further push the boundaries of accuracy and generalization. The process of manually identifying optimal features for machine learning can sometimes limit the detection of complex signal patterns intrinsic to the data. In contrast, deep learning models, such as convolutional neural networks (CNNs) and recurrent neural networks (RNNs), have demonstrated the capability to autonomously learn directly from raw data across different applications. Considering the sequential and dynamic nature of DUS signals, deep learning methods may excel at discerning subtle features and patterns. This potential has been highlighted in prior research on estimating gestational age from DUS, where deep neural networks demonstrated impressive performance with an average error of less than 0.8 months from data with a 1-month resolution  \citep{katebi23}. Moreover, as the volume of DUS data continues to grow, the scalability of deep learning models offers a distinct advantage. Another notable advantage of deep learning frameworks, such as TensorFlow, is their adaptability to edge computing systems such as the mobile phone. 

Building on these insights, this study aims to develop a deep learning for 1D-DUS signal quality assessment classifier that can continue to learn from new data, and can run in real-time on TPU-enabled smartphone running TensorFlow Lite. By integrating this enhanced classification method into the signal extraction system, we intend to provide real-time feedback to TIMs during signal recording. \add{This real-time feedback not only is likely to lead to much improved data quality but also acts as an instantaneous educational tool for TIMs, guiding them on the improved positioning of the 1D-Doppler device to obtain higher quality signals, thus increasing the information extracted from 1D-DUS devices and their utility, ultimately contributing to better monitoring and thereby improving fetal-maternal health. A higher quality recorded signal is likely to increase the information extracted from 1D-DUS devices and their utility, ultimately contributing to better monitoring and thereby improving fetal-maternal health.}

\section{Methods}
\subsection{Data Collection}
The data used in this study originated from a randomized controlled trial in Tecpán, Chimaltenango, a rural highland region of Guatemala.
The trial was approved by the Institutional Review Boards of Emory University, Wuqu' Kawoq $\|$ Maya Health Alliance, and Agnes Scott College (Reference numbers IRB00076231, 2015 001 and 02.02.2015 respectively) and registered on ClinicalTrials.gov (identifier NCT02348840). More details on the design and implementation of the data collection system, and the training of the traditional birth attendants can be found in \citet{stroux16}, \citet{martinez17}, and  \citet{Martinez2018}.

During the trial, 19 TIMs utilized a handheld 1D-DUS device (AngelSounds Fetal Doppler JPD-100s, Jumper Medical Co., Ltd., Shenzhen, China) with a transmission frequency of $3.3\,\mathrm{MHz}$. The device was connected to a smartphone via a bespoke audio cable modified with a capacitor-resistor network to simulate the action of plugging in a hands-free kit to the device. Data were saved as 16-bit uncompressed WAV files with a 44.1\,kHz digitization sampling frequency. Additionally, the DUS device was connected to a speaker, which enabled real-time audio checks by the TIMs and helped to verify the quality of Doppler signals by listening to the heartbeats of the fetus. A bespoke app on the low-cost smartphones guided the TIMs through a range of tasks. This application not only compiled crucial medical data and captured ultrasound recordings but also integrated an alert system for serious concerns, prompting immediate contact with healthcare personnel. It also streamlined data uploads to Amazon Web Services via SMS, GPRS, and WiFi. To effectively use this setup, TIMs participated in four half-day training sessions, followed by an exam to assess proficiency.

The data used in this study includes 191 DUS signals recorded from 142 singleton pregnancies, all of Indigenous Maya women in their second to third trimesters. The median duration of the recordings was found to be 10.2 minutes, with an interquartile range of 1.3 minutes. A histogram of the recording durations is presented in Figure~\ref{fig:Duration_hist}. More detailed information on the recordings can be found in \citet{valderrama18}.

\begin{figure}[htbp]
\floatconts
  {fig:Duration_hist}
  {\caption{Histogram of the DUS recording duration. Note that the signal was broken into 5 minute segments (in case of file corruption) and later reconstituted into single file. This explains the periodicity in the distribution.}}
  {\includegraphics[width=1\linewidth]{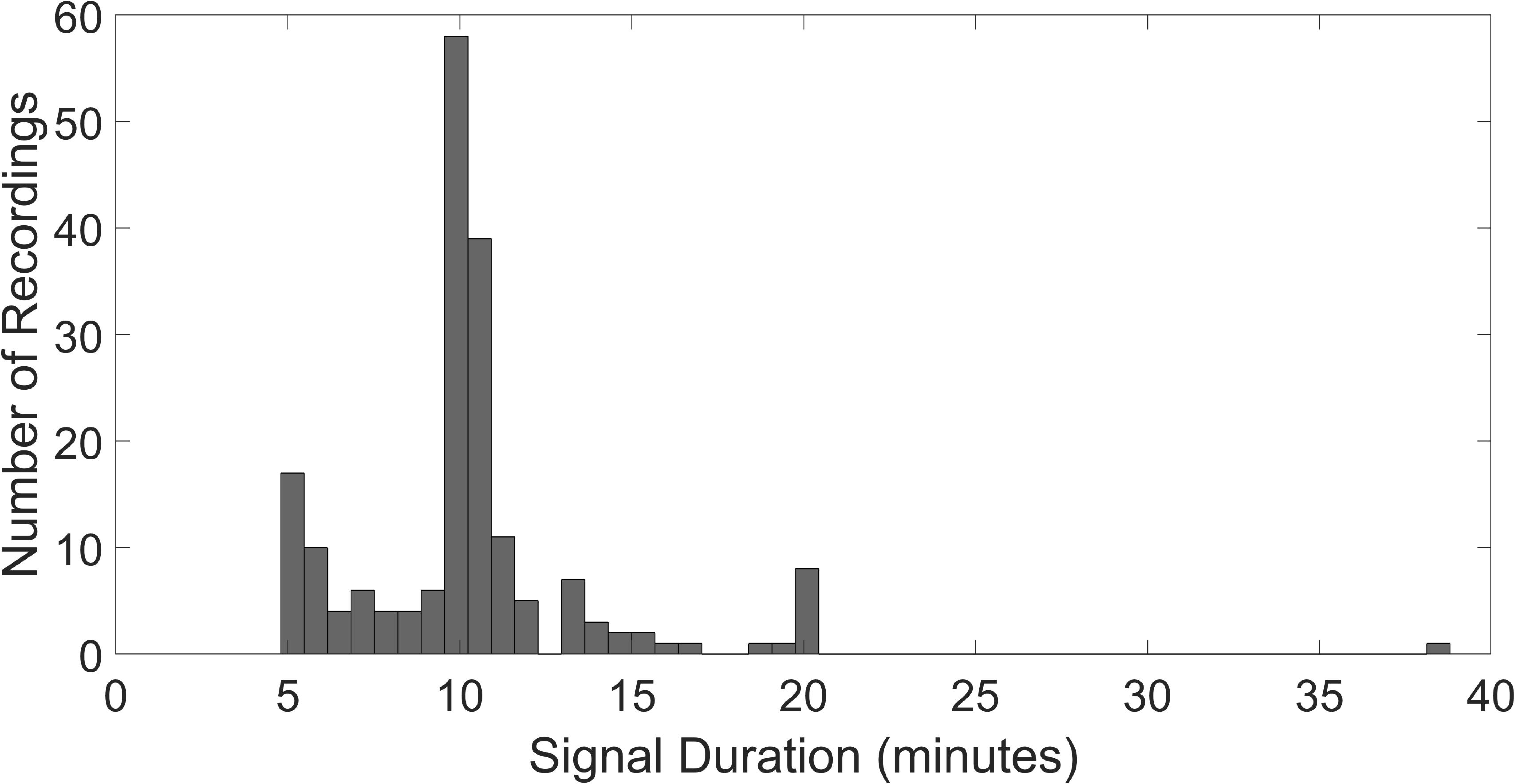}}
\end{figure}

\subsection{Annotation of DUS Recordings}
\label{sec:annotation}
Each DUS recording was decimated to $4\,\mathrm{kHz}$ prior to annotation. For the annotation process, three independent working engineers, familiar with the data, audibly and visually evaluated all of the 191 DUS recordings using a bespoke graphical user interface written in MATLAB (MathWorks, Natick, MA, USA), as described in \citet{valderrama18}. This interface subdivided the recordings into segments lasting 0.75\,s each. Following this segmentation, annotators categorized each segment into one of six distinct classifications:

\begin{itemize}
    \item \textbf{Interference}:  Dominated by electrical disturbances, often manifesting as buzzing noises. These generally came from other mobile phones.
    \item \textbf{Silent}: Segments that were either barely audible or entirely silent, largely due to the audio cable being only partially inserted into the Doppler device.
    \item \textbf{Talking}: Segments potentially containing audible heartbeats but also featuring human speech or environmental sounds, such as animal noises. This was largely due to the audio cable being only partially inserted into the phone, thereby not triggering the external-to-internal microphone switch-over. 
    \item \textbf{Poor}: Noises that did not align with the characteristics of the other defined categories. These can be due to movement of the mother, fetus or TIM, or just poor siting of the Doppler device.
    \item \textbf{Good}: Segments dominated by clear audible heartbeats devoid of the presence of any categories mentioned above.
    \item \textbf{Unsure}: Segments containing a mix of sounds or those that annotators found challenging to place in a definitive category.
\end{itemize}

\subsection{Data Preparation}
In concordance with the previous studies, signal quality in this research was categorized using $3.75\,\mathrm{s}$ windows \citep{Valderrama17, valderrama18}. (3.75\,s is the standard window for heart-rate estimation in cardiotocogrpahy, with a good trade-off between sufficient data and stationarity \cite{valderrama19}.) Each of these windows was composed of five consecutive 0.75\,s signal segments (individually annotated). There was a noted variance in the categorization of the $0.75\,\mathrm{s}$ segments among the three annotators. Therefore, segments were only incorporated into the `Good', `Poor', and `Silent' categories if there was unanimous agreement among all annotators. \add{The total segments for the `Good’, `Poor’, and `Silent’ categories were 26,103, 14,796, and 18,588, respectively.} However, for the `Interference' and `Talking' categories, due to a limited number of segments, those that were labeled similarly by at least two annotators were utilized. \add{For the `Interference’ category, 655 segments were agreed upon by all annotators, and 2,178 were agreed upon by two. For the `Talking’ category, the numbers are 5,043 and 3,682 segments agreed upon by three and two annotators, respectively.} Any segments that were categorized as `Unsure' were subsequently removed from the analysis. \add{A total of 23,929 segments were annotated as `Unsure' by at least two annotators.} 

Finally, $3.75\,\mathrm{s}$ windows were formed by combining five consecutive segments of the same class. To ensure optimal data utilization, a $0.75\,\mathrm{s}$ segment could be repeated in various $3.75\,\mathrm{s}$ windows, leading to potential overlaps between these windows.\add{By incorporating this overlap, it is ensured that the maximum information is extracted from the data, especially from the boundaries of each segment, which might otherwise be overlooked or underrepresented.} This approach was similar to that used in the previous study \citep{valderrama18}. Table~\ref{tab:window-number} presents the distribution of the available $3.75\,\mathrm{s}$ windows for each class.

\begin{table}[tb]
\floatconts
  {tab:window-number}
  {\caption{Number of $3.75\,\mathrm{s}$ windows per each class}}%
  {%
    \begin{tabular}{lc}
    \hline
    \abovestrut{2.2ex}\bfseries Class & \bfseries Number of Windows\\\hline
    \abovestrut{2.2ex}Good & 22,008\\
    Poor & 9,874\\
    Interference & 1,626\\
    Talking & 5,975\\
    \belowstrut{0.2ex}Silent & 17,260\\\hline
    \end{tabular}
  }
\end{table}

After extracting segments of a $3.75\,\mathrm{s}$ duration, a second-order band-pass Butterworth filter was applied to mitigate interference from unwanted frequencies. The optimal cutoff frequencies, derived from the cardiac wall velocity and the $3.3\,\mathrm{MHz}$ frequency of the DUS devices, were identified as $25\,\mathrm{Hz}$ and $600\,\mathrm{Hz}$ in prior research \citep{valderrama19}. Following band-pass filtering, a scalogram was generated using the Morlet mother wavelet. This scalogram offers a two-dimensional representation of a signal, illustrating the progression of its frequency components over time. The scalogram was then normalized using its minimum and maximum values. This normalized scalogram served as the input for the deep learning model, which is discussed in Section~\ref{sec:network}.

\subsection{Network Architecture}
\label{sec:network}
\add{The deep neural network model implemented in this study was influenced by a previous model employed for estimating gestational age from 1D-DUS signals \citep{katebi23}. We adopted the same framework here, as it has been shown to capture physiologically important characteristics of 1D-DUS from non-physiological ones. Moreover, future iterations of the model may be designed for multi-task functions, specifically using a shared framework to simultaneously identify gestational age, signal quality, and fetal heart rate. Finally, we note that by converting the final model to a TFLite model, we enable the code to easily be ported to a mobile phone.}

\add{The 1D-DUS signals naturally encapsulate the cyclic nature of fetal activity. Components such as wall motion, heart valves, and blood flow exhibit varying velocities and intervals. This results in diverse magnitudes of recorded Doppler frequency shifts, with each component occurring at unique time intervals in relation to others. Given this complexity, scalograms are an apt choice to capture both frequency variations and their temporal relationships, serving as the input for our neural network. Our design is structured around three core components, inspired by the beat encoder network explored in earlier research: 1) a feature extractor using a CNN architecture to pinpoint specific frequency-time patterns; 2) an RNN to discern temporal or sequence-based patterns; and 3) an attention mechanism. The attention mechanism, as demonstrated in the prior study, predominantly focuses on high-quality segments of the signal, offering the potential for effective signal quality classification \citep{katebi23}.}

The details of the proposed network are displayed in Figure~\ref{fig:Arc}. In this model, scalograms derived from $3.75\,\mathrm{s}$ signal windows, having dimensions of 250 samples in time and 40 in frequency, were processed using a CNN. The initial architecture includes a 2D convolutional layer with 32 $3\times3$ filters, activated by the ReLU function, followed by batch normalization, $2\times2$ max-pooling, and a 25\% dropout. This pattern, with the convolutional filters doubling from 32 to 64 and then 128, is repeated in subsequent blocks. In later stages, only the frequency dimension is halved using $1\times2$ max-pooling, preserving the temporal features. 

\begin{figure*}[htbp]
\floatconts
  {fig:Arc}
  {\caption{Architecture of the deep learning model consisting of CNN and GRU layers, followed by an attention mechanism.}}
  {\includegraphics[width=0.9\textwidth]{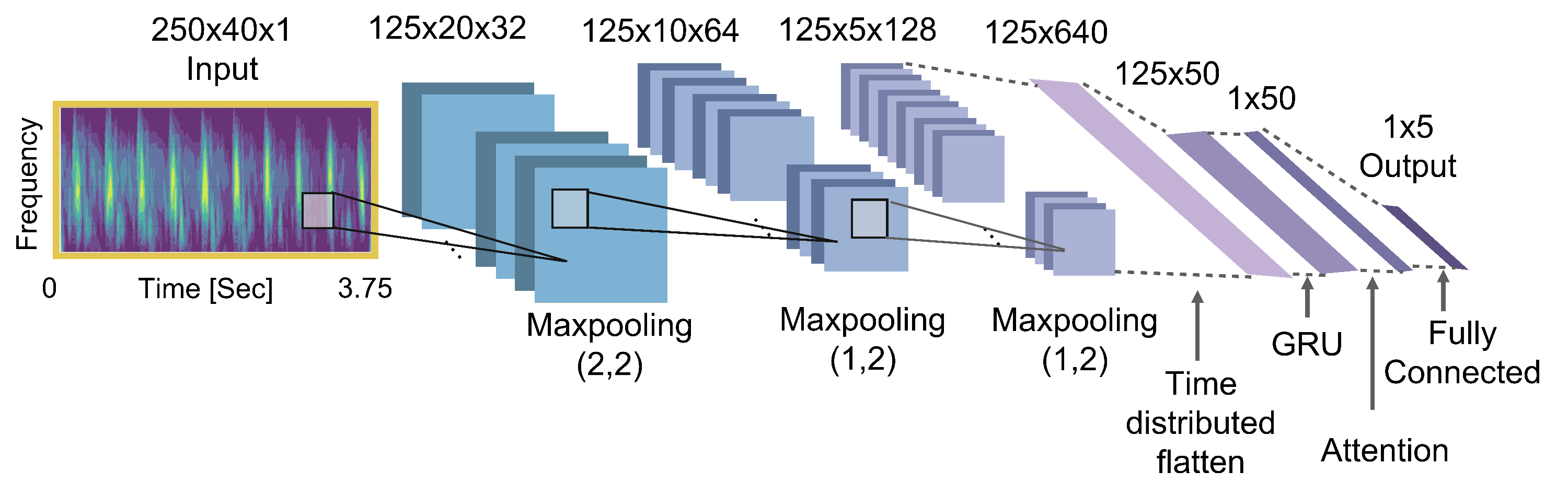}}
\end{figure*}

Subsequent to the convolutional blocks, a TimeDistributed layer is used to flatten the output for sequential processing. A Gated Recurrent Unit (GRU) with 50 units then processes these sequences. The GRU uses two gating mechanisms, namely the reset gate $r$ and the update gate $z$ to control the flow of information. Given the input at time \(t\), \(x_t\), the mathematical formulations for these operations in the GRU can be expressed as:

\begin{equation}\label{eq:GRU}
\begin{aligned}
z_t &= \sigma(W_z x_t + U_z h_{t-1} + b_z), \\
r_t &= \sigma(W_r x_t + U_r h_{t-1} + b_r), \\
\tilde{h}_t &= \tanh(W_h x_t + U_h (r_t \odot h_{t-1}) + b_h), \\
h_t &= (1 - z_t) \odot h_{t-1} + z_t \odot \tilde{h}_t,
\end{aligned}
\end{equation}
Where \( h \) represents the hidden state of the GRU, which retains the memory of past inputs and influences future inferences. The symbol \(\sigma\) represents the sigmoid activation function, and \(\odot\) indicates element-wise multiplication. The parameters \(W\), \(U\), and \(b\) are the weight matrices and bias vectors, respectively, that are specific to each operation.

Following the GRU, a dense layer with 50 units, activated by ReLU, processes the data. An attention network is then utilized, offering attention mechanisms to emphasize significant part of the sequence. The equations governing this attention mechanism are given by:

\begin{equation}\label{eq:attention}
\begin{aligned}
u_t &= \tanh(W h_t + b), \\
\alpha_t &= \frac{\exp(u_t^T u)}{\sum_t \exp(u_t^T u)}, \\
v &= \sum_t \alpha_t h_t.
\end{aligned}
\end{equation}

In the mechanism, the hidden representation at time \( t \), denoted as \( h_t \), undergoes a non-linear transformation to produce \( u_t \). This \( u_t \) is compared with a reference vector \( u \) using the dot product to assess segment relevance. From this, the attention weights \( \alpha_t \) are derived via a softmax function. Consequently, the vector \( v \) is computed as a weighted sum of the \( h_t \) states based on these attention weights. This vector \( v \)  provides a comprehensive representation, encapsulating the essence of the 1D-DUS input signal.

Finally, the model incorporates a dense layer with 5 units activated by a softmax function, classifying data into the five classes described in Section~\ref{sec:annotation}. 

\subsection{Model Implementation and Evaluation}
Based on the network presented in Section~\ref{sec:network}, we assessed the performance using three distinct models. In the first model, referred to as CNN+Att, only the CNN network combined with an attention layer was utilized. The second model, GRU+Att, employed solely the GRU layer integrated with the attention mechanism. In the final model, CNN+GRU+Att, we incorporated all components, consisting of the CNN and GRU networks, along with the attention mechanism. This comparative analysis aids in discerning the most influential components of the network and identifying the most streamlined model yielding optimal results.

For the evaluation of each network, we applied a stratified five-fold cross-validation. The stratification was based on the individual recordings, ensuring each recording appeared in only a single fold, thus preventing information leakage of the same recording between training and test sets. (Here we assume noise is independent between recordings, which is a good first-order approximation as they are taken on different days.) To guarantee the presence of all labels within each fold, the mode of segment classes for each recording was determined, and stratification was then executed based on these modes.

The networks were developed using TensorFlow 2.2.0 and Python 2.7.5. The computational system, equipped with 64\,GB of RAM, a single CPU, and an NVidia Tesla P100 GPU, served for both training and testing the model. The categorical cross-entropy was selected as the loss function, and mini-batch stochastic gradient descent was employed to fine-tune the network parameters. Batch sizes of 128 were utilized. Given the imbalanced distribution of class labels, as presented in Table~\ref{tab:window-number}, the classification accuracy was enhanced by generating balanced batches through the random oversampling of minority classes.

The performance of each model, once trained for each fold, was assessed by the average and standard deviation of precision, recall, and F1 score for every quality class over the 5 folds. In addition, the values of micro and macro F1 scores were computed. The micro F1 score provides a comprehensive measure of a model's performance across all categories. This is achieved by summing the total number of true positives, false positives, and false negatives from all categories, and then computing the precision, recall, and, subsequently, the F1 score from this aggregated data. It is calculated as:
\begin{equation}
\text{Micro F1} = \frac{2 \times \text{Micro Precision} \times \text{Micro Recall}}{\text{Micro Precision} + \text{Micro Recall}},
\end{equation}
where:
\begin{align}
&\text{Micro Precision} = \frac{\sum_{i=1}^5 TP_i}{\sum_{i=1}^5 TP_i + \sum_{i=1}^5 FP_i} \\
&\text{Micro Recall} = \frac{\sum_{i=1}^5 TP_i}{\sum_{i=1}^5 TP_i + \sum_{i=1}^5 FN_i}.
\end{align}
Conversely, the macro F1 score computes the F1 score separately for each of the 5 quality categories and then averages them:
\begin{equation}
\text{Macro F1} = \frac{1}{5} \sum_{i=1}^5 \text{F1}_i
\end{equation}
In these equations, \( TP \), \( FP \), and \( FN \) represent true positives, false positives, and false negatives, respectively.

\subsection{Comparison with Prior Work}
\label{sec:Comparision}
After identifying the best-performing network structure, its performance was compared with the results of the classifier model from prior research \citep{valderrama18}. This comparison was particularly intriguing because that model utilized the same annotated dataset and targeted the same quality classes for classification. It is insightful to determine if the deep learning model could offer more accurate classifications.

In the previous approach, \citet{valderrama18} employed a two-step classification process. Initially, a logistic regression model was trained to identify silent segments based on signal variance. Subsequently, a multiclass SVM was employed to categorize the remaining four classes. For this SVM classification, the best 17 signal-extracted features were utilized. Henceforth, this model is known as ``the SVM model'' for the purpose of this study. 

In \citet{valderrama18}, the dataset was divided into two equal subsets: one comprising DUS recordings from 73 subjects for training and the other from 73 subjects for testing. For a rigorous comparison, we aligned our training and testing datasets with those of the SVM model approach. Nonetheless, since 4 of these recordings were unavailable — 3 from the training set and 1 from the testing set — our model was trained on data from 70 subjects and tested on recordings from 72 subjects. Consequently, there were 98 recordings for training and 93 for testing. It's worth noting that a single fetus can be represented in the dataset (at different gestational ages), taken over different visits. While this creates some moderate correlation, it is not significant when considering the noises are generally independent between visits. We also evaluated the SVM model on the same test set and compared the precision, recall, and F1 score of each quality category as classified by both models.

\section{Results}
\subsection{Model Assessment}
Table~\ref{tab:model-comparison} displays the average F1 scores alongside their standard deviations obtained over a 5-fold cross-validation for the five quality classes across three distinct network architectures: CNN+Att, GRU+Att, and CNN+GRU+Att. From the results, the comprehensive CNN+GRU+Att model had higher average F1 scores in the `Good', `Poor', and `Silent' quality classes compared to the other two models. However, the CNN+Att model—comprising CNN and attention mechanisms—achieved superior F1 scores in the `Interference' and `Talking' classes. A potential reason for this could be the limited data available for these classes (as presented in Table~\ref{tab:window-number}), causing the more complex model to overfit the training data and thus negatively impacting accuracy on the test set. Among the models, the GRU+Att (integrating RNN and attention) showed the lowest performance across all categories.

\begin{table}[tb]
\floatconts
  {tab:model-comparison}
  {\caption{Mean F1 scores (in \%) of the three studied models across five categories, assessed via five-fold cross-validation, along with micro and macro averages of the F1 scores (standard deviations provided in parentheses.)}}
  {\resizebox{\linewidth}{!}{%
  \begin{tabular}{lccc}
  \toprule
  & \textbf{CNN+Att} & \textbf{GRU+Att} & \textbf{CNN+GRU+Att} \\
  \midrule
Good & 98.21(0.72) & 97.25(1.08) & \textbf{99.16(0.30)} \\
Poor & 89.12(9.27) & 86.23(7.89) & \textbf{94.11(3.55)} \\
Interference & \textbf{87.54(4.86)} & 65.81(13.35) & 84.22(6.63) \\
Talking & \textbf{96.79(0.98)} & 84.69(6.15) & 94.88(4.76) \\
Silent & 94.44(7.54) & 97.39(3.02) & \textbf{98.79(1.81)} \\
Micro F1 & 95.12 (4.00) & 93.47 (2.78) & \bf 97.41 (1.24) \\
Macro F1 & 93.22 (3.14) & 86.27 (3.40) & \bf 94.23 (2.44) \\
  \bottomrule
  \end{tabular}
  }}
\end{table}

For the subsequent analyses, the best model was identified by comparing the average micro and macro F1 scores obtained from the five-fold cross-validation, as illustrated in Table~\ref{tab:model-comparison}. The CNN+GRU+Att model exhibited the highest performance with average micro and macro F1 scores of 97.41\% and 94.23\%, respectively, indicating its superior classification efficacy. Consequently, the CNN+GRU+Att model was selected as the best model and will be employed for further evaluation.

For a more in-depth insight into the CNN+GRU+Att model's classification capability across various quality categories, Table~\ref{tab:pr-re-f1} provides precision, recall, and F1 score values, averaged over the 5-fold cross-validation. To complement this, a cumulative confusion matrix that aggregates classifications across the five folds is provided in Table~\ref{tab:confusion-matrix}. The `Good' class emerges as the most accurately classified category, with all metrics exceeding 99\%. In contrast, the `Interference' was the most challenging class, with an F1 score of 84\%, which is possibly attributed to its minimal training labels as highlighted in Table~\ref{tab:window-number}. The confusion matrix further illustrates that the erroneously classified `Interference' segments predominantly fall under the `Good', `Poor', and `Talking' categories.
\begin{table}[tb]
\floatconts
  {tab:pr-re-f1}
  {\caption{Precision, Recall, and F1 Score values for the CNN+GRU+Att model, averaged over five-fold cross-validation. Numbers in parentheses indicate standard deviation.}}
  {\resizebox{\linewidth}{!}{%
  \begin{tabular}{lccc}
  \toprule
  \bfseries Category & \bfseries Precision(\%) & \bfseries Recall(\%) & \bfseries F1 Score(\%) \\
  \midrule
  Good & $ 99.17 (0.69)$ & $99.15 (0.37)$ & $99.16 (0.30)$ \\
  Poor & $96.41 (0.77)$ & $92.22 (6.95)$ & $94.11 (3.55)$ \\
  Interference & $88.28 (12.84)$ & $83.21 (11.65)$ & $84.22 (6.63)$ \\
  Talking & $93.98 (8.36)$ & $96.23 (2.23)$ & $94.88 (4.76)$ \\
  Silent & $97.83 (3.32)$ & $99.83 (0.19)$ & $98.79 (1.81)$ \\
  \bottomrule
  \end{tabular}}}
\end{table}
\begin{table}[tb]
\floatconts
{tab:confusion-matrix}
{\caption{Cumulative confusion matrix over five-fold cross-validation for the CNN+GRU+Att model}}
{\resizebox{\linewidth}{!}{%
\begin{tabular}{ccccccc}
\multicolumn{1}{c}{}    & \multicolumn{6}{c}{Estimated Class}                                                                                                                                       \\
\multirow{6}{*}{\rotatebox[origin=c]{90}{Actual Class}} &                                   & Good                   & Poor                   & Interference           & Talking                  & Silent                 \\ \cline{3-7} 
& \multicolumn{1}{r|}{Good}         & \multicolumn{1}{c|}{21831} & \multicolumn{1}{c|}{109} & \multicolumn{1}{c|}{56} & \multicolumn{1}{c|}{12}      & \multicolumn{1}{c|}{0} \\ \cline{3-7} 
& \multicolumn{1}{r|}{Poor}         & \multicolumn{1}{c|}{72} & \multicolumn{1}{c|}{9106} & \multicolumn{1}{c|}{66} & \multicolumn{1}{c|}{297}     & \multicolumn{1}{c|}{333} \\ \cline{3-7} 
& \multicolumn{1}{r|}{Interference} & \multicolumn{1}{c|}{85} & \multicolumn{1}{c|}{89} & \multicolumn{1}{c|}{1396} & \multicolumn{1}{c|}{50}     & \multicolumn{1}{c|}{6} \\ \cline{3-7} 
& \multicolumn{1}{r|}{Talking}      & \multicolumn{1}{c|}{29} & \multicolumn{1}{c|}{113} & \multicolumn{1}{c|}{32} & \multicolumn{1}{c|}{5790}    & \multicolumn{1}{l|}{11} \\ \cline{3-7} 
& \multicolumn{1}{r|}{Silent}       & \multicolumn{1}{c|}{0} & \multicolumn{1}{c|}{24} & \multicolumn{1}{c|}{0} & \multicolumn{1}{c|}{3}       & \multicolumn{1}{c|}{17233} \\ \cline{3-7} 
\end{tabular}
}}
\end{table}
\subsection{Evaluating Deep Learning and SVM on Common Data}
The selected CNN+GRU+Att network, hereafter also referred to as the Deep Learning (DL) model, was trained on the same dataset as the SVM model presented by \citet{valderrama18}, as detailed in Section~\ref{sec:Comparision}. Table~\ref{tab:DL-SVM} compares the precision, recall, and F1 score results of both the DL and SVM models evaluated using an identical test set.
\begin{table}[tb]
\floatconts
{tab:DL-SVM}
{\caption{Comparison of precision, recall, and F1 scores for the Deep Learning (DL) model from the current study and the Support Vector Machine (SVM) from \citet{valderrama18} across five Categories in the test set.}}
{\resizebox{\linewidth}{!}{%
{\begin{tabular}{lcccccc}
\toprule
\multirow{2}{*}{Category} & \multicolumn{2}{c}{Precision(\%)} & \multicolumn{2}{c}{Recall(\%)} & \multicolumn{2}{c}{F1 Score(\%)} \\
\cmidrule(r){2-3} \cmidrule(r){4-5} \cmidrule(r){6-7}
& DL & SVM & DL & SVM & DL & SVM \\
\midrule
Good & \bf99.36 & 97.06 & \bf98.79 & 93.97 & \bf99.07 & 95.49 \\
Poor & \bf95.21 & 86.05 & \bf94.15 & 91.42 & \bf94.68 & 88.65 \\
Interference & 64.40 & \bf85.13 & 81.44 & \bf92.44 & 71.93 & \bf88.63 \\
Talking & \bf98.17 & 95.87 & 93.69 & \bf96.63 & 95.88 & \bf96.25 \\
Silent & 96.84 & \bf100.00 & \bf99.98 & 99.91 & 98.39 & \bf99.96 \\
\bottomrule
\end{tabular}}}}
\end{table}
The DL model surpassed the SVM model in all metrics for the `Good' and `Poor' classes. Notably, the classification performance for the `Good' category was especially very high, with an F1 score exceeding 99\%. However, the SVM model exhibited better capability in categorizing signals with `Interference'. The performance metrics for `Talking' and `Silent' categories were close for both models, but the SVM marginally outperformed the DL model in terms of F1 score.

\section{Discussion}
\subsection{Performance and Limitations of the CNN+GRU+Att Model}
The CNN+GRU+Att model demonstrated its robustness in comparison to the simpler architectures, such as CNN+Att and GRU+Att. One notable strength of the CNN+GRU+Att model is its ability to capture both local features, via the CNN layers, and temporal dependencies through the GRU layers. This simultaneous feature extraction likely contributed to its superior performance in the `Good', `Poor', and `Silent' classes. Conversely, the GRU+Att model had the lowest performance across all classes. This highlights that solely focusing on temporal relationships may be inadequate. The inclusion of CNN layers appears essential for identifying patterns within the signals. 

Nevertheless, the CNN+GRU+Att model struggled to outperform the simpler CNN+Att model in the `Interference' and `Talking' classes. This implies that model complexity, while beneficial, can pose risks, especially when training data is scarce, leading to potential overfitting. It is worth noting that the labeled data for the `Interference' and `Talking' categories might also be noisier (with lower inter-rater agreement levels) than the other three. As described in Section~\ref{sec:annotation}, the annotations for the `Good', `Poor', and `Silent' classes were used under conditions where all three annotators were in agreement. Yet, due to data scarcity for `Interference' and `Talking', labels were considered even with the consensus of only two annotators. This observation emphasizes the vital importance of having robust and sufficient training data for deep learning models to achieve effective generalization.

\subsection{Deep Learning vs Support Vector Machine in Quality Classification}
The comparison between the DL model and the SVM model of \citet{valderrama18}, highlights the advantages of the DL approach. Notably, the DL model achieved an F1 score of 99.07\% for the `Good' class, outperforming the SVM model's 95.45\%, as detailed in Table~\ref{tab:DL-SVM}. It is essential to accurately detect these high-quality DUS signal segments, as they are the foundation for further analyses and are vital for effective fetal health monitoring. While recognizing unwanted signal categories, such as `Interference', `Talking', and `Silent', is informative for the TIMs during signal recording, the primary focus remains on differentiating between good and low-quality signals. Adjustments in hardware configurations can mitigate issues like 'Interference', 'Talking', and 'Silent'. However, low-quality data resulting from suboptimal Doppler usage presents a more challenging obstacle. This challenge underscores the primary objective of this study: offering AI-assisted training to navigate these issues. This further emphasizes the advantage of the DL model for signal quality classification.

Additionally, it is important to note that, although the evaluation metrics associated with the DL model for the classification of 'Silent' DUS segments appear lower, the model might still excel under general conditions. The SVM model employs a simple logistic regression based on signal variance to identify silent signals. In this method, any signal with variance below a set threshold is labeled as `Silent'. However, this threshold may not be consistent when the DUS signal is obtained using different devices or in varied application settings. The DL model, with its capacity to identify patterns and relationships in the raw signal, offers a more adaptable and universally relevant classification mechanism. This flexibility enhances its potential for a wide variety of recording environments.

\subsection{Implications and Future Directions}


\add{Having demonstrated the capabilities of the model introduced in this study, there is significant potential for its integration into our system for real-time quality assessment. Utilizing the TFLite version of the deep learning algorithm promises to be more efficient and adaptable than relying on a series of feature extraction algorithms that feed into an SVM. We have already developed an Android app prototype using TFLite, which has shown encouraging initial results. The next step involves comprehensive field evaluations to determine the model's effectiveness and applicability in real-world settings. Furthermore, our designed architecture is compatible with other deep learning-based algorithms we have developed, such as those for estimating gestational age and blood pressure. This provides a platform for the creation of a multi-task algorithm, which will be the focus of our future work.}

Furthermore, due to the scalability of deep learning models, enhanced performance is anticipated as more annotated data is collected, particularly for the 'Interference' class with the lowest accuracy. In this study, a clear correlation was observed between the quantity of available data in each class and the model's performance for that class. Better results were yielded by classes with a larger number of labels. For example, the highest evaluation metrics, as presented in Table~\ref{tab:pr-re-f1}, were exhibited by the 'Good' and 'Silent' classes, which had the greatest number of labeled data (as referenced in Table~\ref{tab:window-number}). Additionally, when comparing the outcomes of a model trained on a smaller dataset, there was a more pronounced performance dip for classes with fewer data samples. A comparison between Tables~\ref{tab:DL-SVM} and \ref{tab:pr-re-f1} reveals that the F1 score for interference dropped from 84.22\% to 71.93\%. Consequently, enhanced model performance is expected by gathering more data, especially for classes with fewer samples, such as 'Interference' and 'Talking'. Future research will focus on training the model using a broader collection of recordings.

\add{Finally, the proposed model, for real-time 1D-DUS signal quality assessment introduces two iterative learning schemes. The first involves incremental retraining of the model in the light of new data. The user can label data in real-time, primarily through objecting to the classifier's output. This label is then saved with the raw data for later relearning. This approach also ensures the model's adaptability across diverse data collection settings or environments. The other form of retraining is from the machine learning model to the user. Over time, the user will become more adept at perceiving the audible qualities of good quality data, and become better at 'locking on' to an improved signal.}

\section{Conclusions}
In this study, a deep learning model was presented for the classification of the quality of 1D-DUS recordings. These recordings were captured by TIMs with minimal training via an affordable Doppler device connected to a smartphone. Specifically, the model was found capable of classifying $3.75\,\mathrm{s}$ segments of the signals into five categories with high accuracy. The accuracy of this model surpassed that of conventional machine learning algorithms previously used, particularly in the classification of `Good' quality signals, achieving an F1 score of over 99\%. This algorithm provides a reliable tool for integration into smartphone apps used by TIMs, enabling real-time feedback on the quality of recorded data. This ensures that better data will be available for future analysis, aiding in the improvement of fetal health in affordable settings. Finally, there exists potential for further refinement of this model through the acquisition and annotation of additional DUS signals.

\acks{Acknowledgements}
This work was supported by a Google.org AI for the Global Goals Impact Challenge Award, the National Institutes of Health, the Fogarty International Center and the Eunice Kennedy Shriver National Institute of Child Health and Human Development, grant \#'s 1R21HD084114 and 1R01HD110480. NK is funded by a PREHS-SEED award grant \# K12ESO33593.

\bibliography{references}

\end{document}